# Strong Coupling of Light to Collective Terahertz Vibrations in Organic Materials

R. Damari[1], O. Weinberg[1], N. Demina, D. Krotkov, K. Akulov, A. Golombek,

T. Schwartz* and S. Fleischer*

School of Chemistry, Raymond and Beverly Sackler Faculty of Exact Sciences

and Tel Aviv University Center for Light-Matter Interaction, Tel Aviv University,

Tel Aviv 6997801, Israel.

*Corresponding authors: talschwartz@tau.ac.il, sharlyf@post.tau.ac.il

[1]These authors contributed equally to this work

**Several years ago, it was shown that strong coupling between an electronic transition in organic molecules and a resonant photonic structure can modify the electronic landscape of the molecules and affect their chemical behavior. Since then, this new concept has evolved into a new field known as polaritonic chemistry, which employs strong coupling as a new tool for controlling material properties and molecular chemistry. An important ingredient in the progress of this field was the recent demonstration of strong coupling of molecular vibrations to mid-infrared resonators, which enabled the modification of chemical processes occurring at the electronic ground-state of materials. Here we demonstrate for the first time strong coupling with collective, intermolecular vibrations occurring in organic materials in the Terahertz frequency region. Using a tunable, open-cavity geometry, we measure the temporal evolution and observe coherent Rabi oscillations, corresponding to a splitting of 68 GHz and approaching the ultra-strong coupling regime. These results take strong light-matter coupling into a new class of materials, including polymers, proteins and other organic materials, in which collective, spatially extended degrees of freedom participate in the dynamics.**

When light is compressed into a region comparable to its wavelength, its interaction with matter can overcome all the incoherent and dissipative processes, which profoundly changes its nature. In this regime, known as strong coupling, the wavefunctions of the photons and the material excitations are coherently mixed to form cavity polaritons, which are hybrid light-matter quantum states[1]. This fascinating phenomenon has been observed in many different types of material systems, such as cold atoms[2–4], excitons in semiconductors[5,6], electronic spins in nitrogen-vacancy centers[7], phonons in inorganic crystals[8–12] and many others. Among these, strong coupling with organic molecules[13,14] has seen an ever-increasing interest in recent years, both in conventional Fabry-Perot microcavity systems as well as in plasmonic structures[15]. Interestingly, the creation of the polaritonic wavefunctions under strong coupling and the modification of the energetic landscape of the molecules can have a significant influence on the physical and chemical properties of the molecules[1,16,17], affecting the rates and yields of chemical reactions[18–23], their emission properties[24–26], electronic and excitonic transport[27–31] and more. This new field, known as polaritonic chemistry, is currently under intense study, both experimentally and theoretically. While traditionally organic strongly-coupled systems involved the coupling of an optical resonance to electronic transitions in molecules (Frenkel excitons), recently, vibrational strong coupling has been introduced as a new paradigm[32–37]. In such systems, a particular intramolecular, optically-active vibrational transition is coupled to a mid-infrared resonator, creating hybrid excitations termed "vibro-polaritons". As has been demonstrated over the past few years, the creation of such vibro-polaritons allows the manipulation of molecular processes occurring at the electronic ground-state, by targeting a specific bond in the molecules[19,38,39].

Here we demonstrate, for the first time, strong coupling of collective vibrations in ensembles of organic α-lactose molecules, occurring at THz frequencies ($10^{11}$-$10^{13}$ Hz). Unlike the previously studied vibrational strong coupling, here the cavity mode is coupled to inter-molecular vibrations in molecular crystallites, i.e. oscillatory motion of the hydrogen-bonded molecules with respect to one another. Interestingly, we observe the Rabi-splitting typical of strong coupling and coherent Rabi-oscillations at room temperature, despite the fact that the energy of the collective vibrational transition ($\hbar \nu_{vib}$~2 meV), as well as the light-matter interaction strength are much lower than $k_BT$ (~25 meV).

Our results extend the applicability of polaritonic chemistry to other organic large-scale systems, such as biological macromolecules[40], polymer chains[41] and energetic materials with low lying collective vibrations[42].

Lactose, which is found in milk, is a disaccharide composed of galactose and glucose. In this study we use α-lactose monohydrate, which is one of the anomers formed upon the crystallization of lactose, with the chemical structure shown in Fig. 1a. The α-lactose powder used in this study (Sigma-Aldrich) is comprised of small, polycrystalline particles, a few tens of microns in size, as shown in Fig. 1b. To measure its THz absorption spectrum, we prepared a ~1.3 mm-thick pellet of pristine α-lactose using a pressing die (see Methods Section), and measured its absorption spectrum using terahertz time-domain spectroscopy (THz-TDS). The result is presented in Fig. 1c, showing a sharp absorption peak at 0.53 THz (17.7 $cm^{-1}$) and a width of 21 GHz FWHM. This absorption line corresponds to a collective, intermolecular vibration in the molecular crystal, in which the molecules move with respect to each other as a rigid body[43–45]. An additional weaker absorption peak is observed within our usable THz bandwidth, at 1.2 THz. In order to demonstrate strong coupling of the collective vibrational mode at 0.53 THz, we utilized the open-cavity geometry[46] depicted in Fig. 1d (see Method Section for further details). The cavity is composed of two Au mirrors prepared by sputtering ~6 nm Au layers on 1 mm-thick quartz substrates. The reflection amplitude of the mirrors was measured to be ~90% for the THz field (81% reflectivity). In the open cavity geometry, one of the mirrors is fixed, while the other is mounted on a computer-controlled translation stage, parallel to the fixed mirror, such that the cavity length $d$ can be varied continuously.

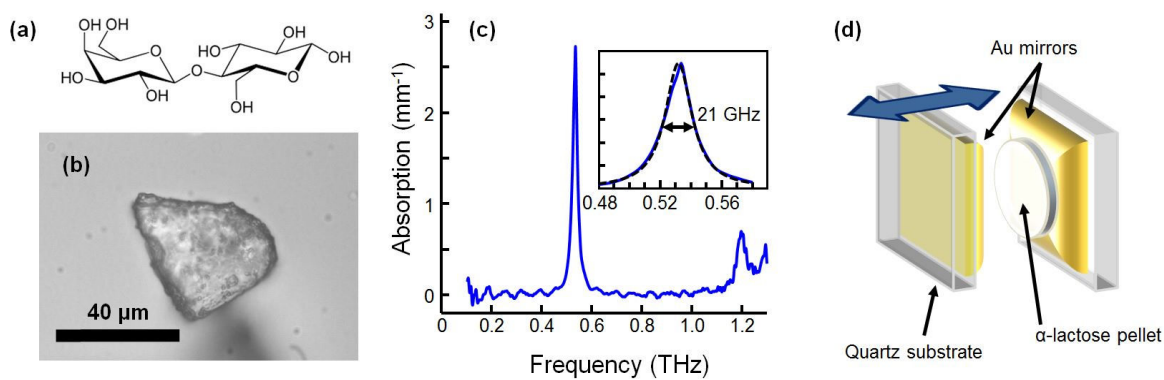

**Fig. 1.** (**a**) Chemical structure of α-lactose. (**b**) A microscope image of an α-lactose crystallite. (**c**) Measured THz absorption spectrum of crystalline α-lactose pellet. The inset shows the fit of the measured absorption peak (blue line) to a lorentzian line-shape (black dashed line). (**d**) A sketch of the open THz microcavity used in the experiments.

The measurements were performed using a home-built, time-domain terahertz spectrometer, shown in the schematic diagram in Fig. 2a. In a typical measurement, an ultrashort laser pulse (100fs pulse duration, 800nm central wavelength) from a Ti:Sapphire chirped pulse amplifier (Legend Duo, Coherent Inc.) is split to form a strong optical beam for THz generation and a weak readout pulse for time-resolved electro-optic sampling of the THz field[47,48]. A single-cycle THz pulse is generated via tilted pulse-front optical rectification in $LiNbO_3$ (LN)[49] and focused through the sample (S), which is placed at the focal plane of a 4-f setup composed of two off-axis parabolic reflectors. The THz field and the readout pulse are combined by a pellicle beam-splitter (PBS) and focused at the electro-optic detection crystal (Gallium Phosphate, GaP), following which the probe beam is analyzed for its differential polarization changes (by splitting the two polarizations with a Wollaston Prism [WP] and a pair of photodiodes [PD]). The signals are detected using a lock-in amplifier (Stanford Research Systems, SR830) and recorded by a desktop PC. The entire system is purged with dry air (relative humidity < 4%) to eliminate THz absorption by the water vapor in the ambient lab atmosphere (water absorption lines of 0.56 THz, 0.75 THz and 0.98 THz are clearly observed in our THz spectrum and are completely removed when the system is purged with dry air). Also shown in Fig. 2 are the measured electric field of the generated THz pulse (Fig. 2b) and its calculated power spectrum (Fig. 2c). As seen, the input field is indeed a single-cycle pulse, centered around 0.6 THz with a usable spectrum covering the 0.1-1.2 THz range.

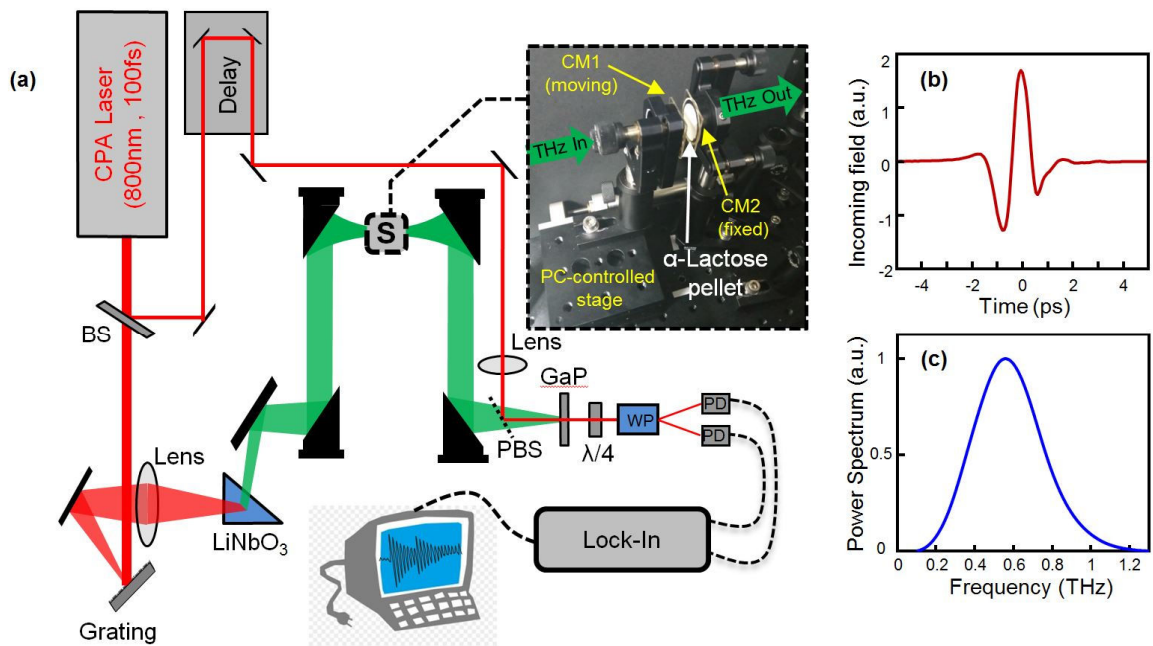

**Fig. 2.** Time-domain THz spectrometer. (**a**) Schematic sketch of the optical setup. The inset shows a photograph of the open cavity, which is formed by a moveable mirror (CM1) and a fixed mirror (CM2). (**b**) The time-resolved single-cycle THz field used in this work and (**c**) its power spectrum, obtained by the Fourier transform of the data in Fig. 2b.

First, we characterized the response of the empty cavity, i.e., when the gap between the mirrors contains only dry air. Note that in such time-domain experiments, parasitic reflections within the THz-TDS spectrometer result in spurious signals which enter the measurements. These distortions are removed by deconvoluting the instrument response function from the raw signals[50,51]. The results of the THz-TDS (following deconvolution) are presented in Fig. 3a, showing the time-domain signal of the field exiting the cavity, for several different cavity length values (at normal incidence). As can be seen, when the single-cycle THz pulse passes through the cavity, it is stretched to an exponentially-decaying oscillatory signal, as expected for a resonant cavity with a finite lifetime. The transmission spectrum of the cavity can then be calculated by taking the ratio between the power spectra of the (deconvoluted) transmitted signals shown in Fig. 3a and dividing them by the input pulse power spectrum, shown in Fig. 2c. These transmission spectra, calculated for the different cavity lengths, are shown in Fig. 3b. As can be seen, the resonant Fabry-Pérot cavity modes are clearly visible, with their frequencies obeying the relation $f_m = \frac{c}{2d}m$, where $c$ is the speed of light, $d$ is the cavity length (the distance between the mirrors) and $m$ is the mode number (assigned in Fig. 3b). Specifically, for a cavity length of 640 μm, for which the second-order mode is close to the α-lactose absorption line, the resonant modes have a transmission peak of 0.5-1%, and a linewidth of 14 GHz, matching the calculated Finesse for mirrors with reflectivity of 81%. In addition, we performed T-matrix calculations for the 640 μm cavity to simulate the spectral response of the cavity (solid green line in Fig. 3b), which agree with the experimental measurement.

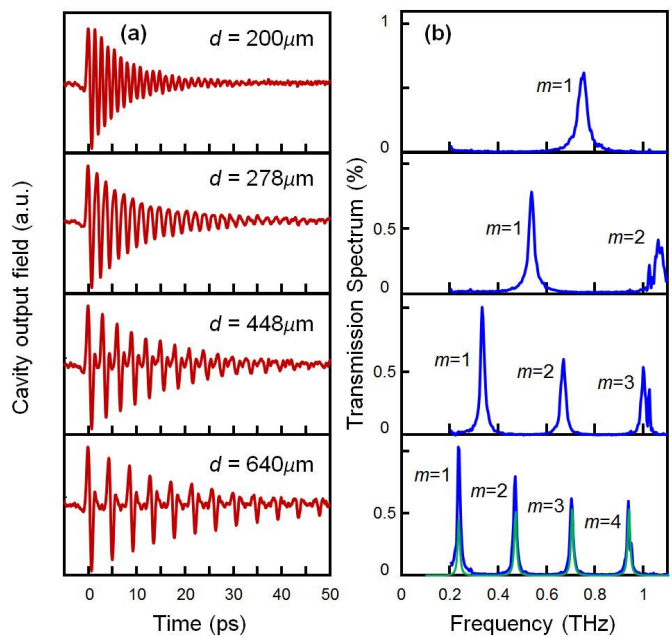

**Fig. 3. Time-domain THz spectroscopy of an empty cavity.** (**a**) Time-resolved THz signal measured at the output of the empty cavity with different distances, *d,* between the mirrors, showing the exponentially-decaying oscillations of the field exiting the cavity. (**b**) Transmission power spectra of the cavity with increasing cavity lengths, showing the gradual progression of the empty-cavity resonant modes. The green line shows the transmission spectrum obtained by T-matrix calculations.

Next, we examined the response of the cavity with the α-lactose pellet placed between the mirrors. We prepared an α-lactose pellet of 250 μm in thickness, attached it to the fixed mirror and adjusted the total cavity length to ~350 μm. Under such conditions, the effective optical length of the cavity (given by $d_{opt} = d_{\alpha L} n_{\alpha L} + d_{air}$ with $d_{\alpha L}$ being the pellet thickness, $n_{\alpha L}$=1.8 the background refractive index of α-lactose[52] and $d_{air}$~100 μm is the thickness of the air-gap) is ~550 μm, such that the second-order cavity mode is resonant with the collective vibrational mode of the α-lactose at 0.53 THz. The time-resolved THz field exiting the cavity is presented in Fig. 4a. We observe a similar exponentially-decaying oscillation, as for the empty cavities, but here the signal is modulated by a periodic envelope. This periodic modulation corresponds to Rabi-oscillations in the cavity, signifying the strong coupling between the collective vibrations of the α-lactose crystallites and the cavity. The transmission spectrum of the α-lactose cavity, obtained using the Fourier transform of the signal in Fig. 4a, is shown in Fig. 4b (blue solid line). Furthermore, by fitting the results to T-matrix calculations (using the experimentally-measured refractive index of α-lactose[52]), shown by the green solid line, we obtain a thickness of $d_{\alpha L}$=250 μm for the α-lactose pellet and $d_{air}$=97 μm for the air gap thickness. As can be seen in both the experimental measurement and the simulations, the hybrid cavity/α-lactose system exhibits a clear splitting in the spectral response around the collective vibration frequency and the formation of two THz vibro-polariton states, at 0.50 and 0.56 THz, indicating once again that the hybrid system is indeed within the strong coupling regime. In addition to the polaritonic modes, the first (*m*=1) and third (*m*=3) order cavity modes at 0.25 and 0.83 THz are also located within the bandwidth of the input pulse. Therefore, the single-cycle pulse excites a coherent superposition of the polaritonic modes as well as the non-coupled cavity modes, which gives rise to the seemingly-irregular dynamics seen in the time domain (Fig. 4a). To illustrate this, we numerically filter the time-domain signal by a band-pass filter, leaving only frequencies within the range of 0.38-0.68 THz. The filtered time-domain signal is presented in Fig. 4c. As seen, the Rabi-oscillations of the coupled system are clearly observed, demonstrating the reversible and coherent light-matter interaction taking place in the system. Interestingly, Rabi oscillations occurring under strong coupling of molecular excitons and plasmonic structures were previously observed by probing the excited state population, using ultrafast pump-probe spectroscopy[53]. However, here we are able to observe the Rabi-oscillations in the emitted field directly, including its oscillating phase.

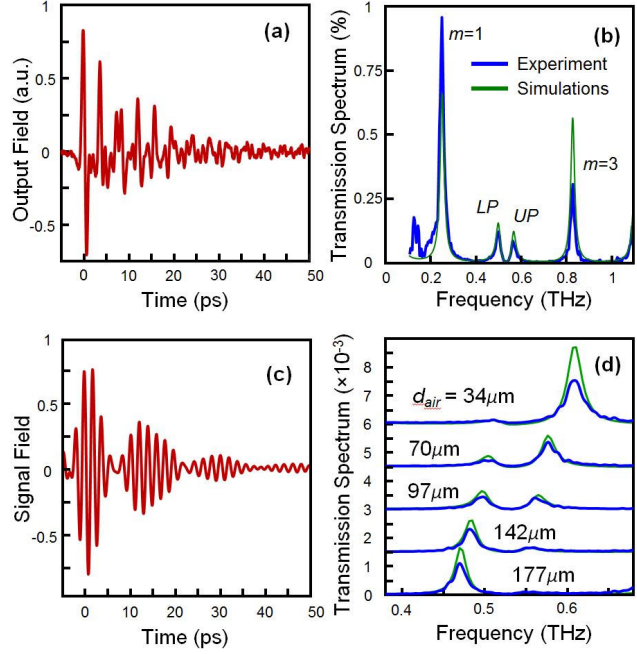

**Fig. 4. Strong coupling of collective THz vibrations in α-lactose.** (a) Measured time-domain output signal from the cavity with the α-lactose pellet, with its length chosen such that the 2$^{nd}$ cavity resonance ($m = 2$) is at 0.53THz. (b) Transmission spectrum of the cavity with α-lactose with a total cavity length of 350 μm, obtained from the experimentally-measured time domain signal in Fig. 4a (blue line) and by T-matrix calculations (green line). The strong coupling between the collective vibrations and the cavity results in a spectral splitting that indicates the formation of hybrid vibro-polariton states. (c) Spectrally-filtered output signal from the cavity, showing the Rabi-oscillations of the coupled system in the time-domain. (d) Selected cavity transmission spectra in the vicinity of the α-lactose absorption peak (0.53 THz), measured with different air gap thicknesses $d_{air}$ and α-lactose pellet thickness of $d_{αL}$=250 μm (blue lines), compared to T-matrix calculations (green lines).

Next, we varied the position of the moveable mirror, repeated the measurement and calculated the spectral response as shown in Fig. 4b, for several different cavity lengths. In Fig. 4d we plot the resulting transmission spectra (blue lines), as well as the simulated T-matrix results (green lines). In our simulations, we fix all the parameters except for the thickness of the air-gap between the pellet surface and the moveable mirror, which is extracted by fitting the simulated transmission to the experimental data. Using these simulations, we can extract the second-order cavity resonance for each value of the cavity thickness, by removing the contribution of the vibrational resonance to the refractive index of the α-lactose, only taking into account its background index of refraction. Finally, we use these results to plot the dispersion of the hybrid molecular/cavity system, i.e. the measured vibro-polariton frequencies as a function of the cavity resonance frequency. As seen in Fig. 5, the dispersion shows the formation of the characteristic polariton branches around the absorption frequency of the α-lactose collective vibration. We fit these measurements to the dispersion resulting from the coupled-oscillator model, given by

$$\nu_\pm = \frac{\nu_c + \nu_{vib}}{2} \pm \frac{1}{2}\sqrt{(\Omega_R/2\pi)^2 + (\nu_c - \nu_{vib})^2 - (\Delta\nu_c - \Delta\nu_{vib})^2} \quad (1)$$

where $\Omega_R$ is the (angular) Rabi frequency, $\nu_{vib} = 0.53$ THz is the collective vibration frequency, $\nu_c$ is the cavity frequency (of the second-order mode), and $\Delta\nu_{vib}=21$ GHz and $\Delta\nu_c=14$ GHz are their linewidths. By fitting Equation (1) to the measured data, we obtain a Rabi frequency value of $\Omega_R/2\pi = 68$ GHz. This value is higher than the decay rates of both the cavity and the uncoupled vibration, confirming that our system is indeed within the coherent, strong coupling regime. Moreover, this value is about 13% of the bare vibration frequency, placing this system close to the ultrastrong coupling regime.

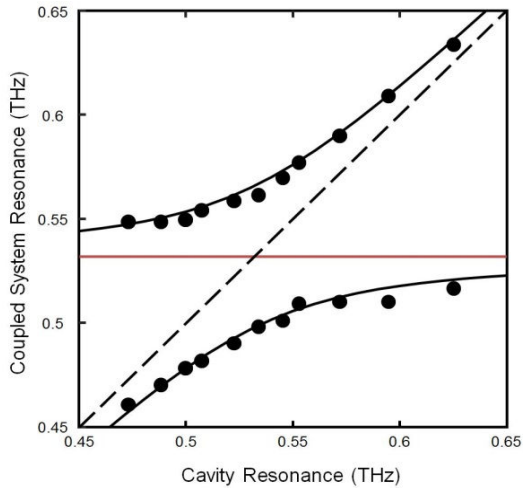

**Fig. 5. THz vibro-polariton branches of strongly-coupled α-lactose.** The circles correspond to the measured polariton peaks, which were obtained by scanning the cavity resonance across the α-lactose absorption (marked by the horizontal red line). The dashed black line marks the empty cavity resonance while the solid black curves show the result of the coupled oscillator model (Eq. 1) fitted to the measured data.

**Discussion**

We have demonstrated the strong-coupling of the collective vibration of α-lactose crystallites and a Fabry-Pérot cavity in the low-THz frequency region (0.53THz), and observed a Rabi-splitting of ~13% of the fundamental frequency. Moreover, we have observed the coherent vacuum Rabi-oscillations taking place in the coupled system, by taking advantage of ability to perform time-domain and phase-sensitive measurements of the THz field. Interestingly, since the measurements are performed at room temperature, the energy of the collective vibration, which is $h\nu_{vib} = 2.2$ meV is lower than $k_B T$. This means that, at steady state, the thermal occupation of the first excited state of the collective vibrational mode is about 48%, placing this system in a very different regime from all previous strongly-coupled organic systems, in which the thermal occupation of the excited state is negligible. Moreover, the Rabi-splitting energy $h\Omega_R = 0.28$ meV is even lower than that, but still, our time-resolved measurements clearly demonstrate coherent Rabi oscillations and coherent light-matter interaction. Finally, in the context of polaritonic chemistry, the ability to affect such low-frequency vibrational modes that extend over several unit cells of the crystal may be used to judiciously control chemical and biological processes that depend on extended degrees of freedom. Affecting the reactivity of energetic materials by modifying their collective intermolecular vibrations (that are also coupled to intramolecular degrees of freedom[54]) or affecting solvent network dynamics that are correlated with protein folding and its kinetics[55] are merely two examples in which material processes can be strongly-coupled to THz cavity modes, in a similar manner to the collective vibrations of α-lactose.

The relatively easy fabrication of the open-mirror-cavity demonstrated here, together with the direct measurement of the amplitude and phase of the electric field in time-domain THz spectroscopy, provide an extended test-bed for studying the very basic underlying physics of the strong-coupling phenomenon. Furthermore, the relatively large (tens to hundreds of μm) length-scale of the THz cavity makes it accessible to additional stimulations, such as optical excitations, that may alter the molecular structure, as well as to structural patterning of the sample to manipulate the light-matter interaction within cavity.

**Methods**

**Open microcavity configuration.** The variable-length open cavity used in this work (see Fig. 2) is composed of a moveable mirror (CM1) and a fixed mirror (CM2). Both mirrors were produced by sputtering a thin layer (6 nm) of gold on a quartz substrate (1 mm thickness), resulting in a transmission amplitude of 90% across the whole usable bandwidth with no apparent spectral dependence. CM1 is mounted on a computer-controlled single-axis stage, with micrometer resolution (<2μm repeatability). By moving CM1 with respect to CM2 we control the length of the cavity and corresponding resonance frequency. CM1 and CM2 are set parallel to each other by coinciding the multiple reflections of a green diode laser from the mirrors at the far field. The α-lactose sample (white, round pellet) was prepared by placing 0.1 g of α-Lactose powder in a pressing die (20mm diameter) at a pressure of 220 kN for 15minutes, which yielded a ∼250μm pellet. The pellet was then glued onto CM2 at a few points around its circumference.

**Transfer matrix calculations.** The simulated transmission spectra of the cavity were calculated using the T-matrix formalism[56]. In these simulations, we used the experimentally-measured refractive index of gold[57] to model the cavity mirrors and adjusted the thickness of the mirrors to match the measured reflectivity of 81%. We note that the fitted thickness of the mirrors was found to be 1.5 nm, which is lower than the actual thickness of 6 nm. This is most probably due to the fact that at such low thicknesses the sputtered metal film is not continuous, but rather composed of small Au islands. The dielectric function of the α-lactose pellet within our usable THz range can be accurately described by a Lorentz-Drude model with contributions from three different vibrational transitions[52]

$$\epsilon_{\alpha L}(\nu) = \epsilon_\infty + \sum \frac{\nu_{p,i}^2}{\nu_i^2 - \nu^2 - i\gamma_i \nu} \qquad (2)$$

where $\epsilon_\infty$=3.2 is the background dielectric constant, $\nu_i$ =0.53, 1.195 and 1.37 THz are the vibrational frequencies, $\gamma_i$=21, 44 and 58 GHz are the linewidths, and $\nu_{p,i}$=0.123, 0.072 and 0.253 are corresponding plasma frequencies.

**Acknowledgments**

This work was supported by the Wolfson Foundation Grant No. PR/ec/20419. S.F acknowledges the support of the Israel Science Foundation (ISF), Grants No. 1065/14, No. 926/18, Grant No. 2797/11 (INREP—Israel National Research Center for Electrochemical Propulsion) and the Wolfson Foundation Grant No. PR/eh/21797. T.S acknowledges the support of the ISF Grant No. 1241/13. K.A. acknowledges the financial support from the Ministry of Science and Technology, Israel.